# Universal Protein Fluctuations in populations of microorganisms

Hanna Salman[1,2], Naama Brenner[3,4*], Chih-kuan Tung[1], Noa Elyahu[3,4], Elad Stolovicki[5,4], Lindsay Moore[5,4], Albert Libchaber[6] & Erez Braun[5,4]

[1]Dept. of Physics and Astronomy, University of Pittsburgh. [2]Dept. of Computational and Systems Biology, School of Medicine, University of Pittsburgh. [3]Dept. of Chemical Engineering, Technion. [4]Laboratory of Network Biology, Technion. [5]Dept. of Physics, Technion. [6]Center for Physics and Biology, Rockefeller University.
[*]Corresponding author.

**The copy number of any protein fluctuates among cells in a population; characterizing and understanding these fluctuations is a fundamental problem in biophysics. We show here that protein distributions measured under a broad range of biological realizations collapse to a single non-Gaussian curve under scaling by the first two moments. Moreover in all experiments the variance is found to depend quadratically on the mean, showing that a single degree of freedom determines the entire distribution. Our results imply that protein fluctuations do not reflect any specific molecular or cellular mechanism, and suggest that some buffering process masks these details and induces universality.**

The protein content of a cell is a primary determinant of its phenotype. However, protein copy number is subject to large cell-to-cell variation even among genetically identical cells grown under uniform conditions ([1-3] and references therein). This variation has been the subject of intensive research in recent years ([4-7] and references therein). Much of this previous work was devoted to characterizing the stochastic properties of various processes underlying gene expression, such as transcription and translation [8], or different stages of the cell cycle [9], and understanding their effect on protein variation. However, gene expression is generally coupled to all aspects of cell physiology, such as growth [10], metabolism [11], aging [12], division [13, 14] and epigenetic processes [15, 16], as well as gene location and function [17], all of which were shown to affect protein variation. The emerging picture is of a plethora of correlated mechanisms at different levels of organization; how they integrate to shape the total protein variation in a dividing population remains an open question [11, 14].

In this work we addressed this question by a phenomenological approach. We measured distributions of highly expressed proteins in proliferating clonal populations of bacteria and yeast under natural conditions, where gene expression is coupled to other cellular processes. By designing an array of different metabolic and regulatory conditions as well as growth environments, we collected a compendium of measurements which systematically covers the major processes of gene expression and cell division, and compared the measured distributions in a wide range of biological realizations. More specifically, our comparisons included: (a) Two archetypical microorganisms, bacteria and yeast, with two well-studied regulatory systems of essential metabolic pathways: the LAC operon in *E. coli* [18] and the GAL system in *S. cerevisiae* [19]. Both systems were studied under environmental conditions in which expression is strongly coupled to metabolism, namely they control the utilization of an essential sugar (lactose and galactose, respectively) as the sole carbon source. (b) Different metabolic growth conditions: the organisms were grown in chemostats – continuous culture in steady state and transients, as well as in batch cultures. (c) Highly regulated versus constitutive expression[1]: the regulated LAC and GAL systems were compared to constitutively expressed proteins in both organisms. (d) Different promoter copy numbers: the same regulatory systems were placed on high-copy (HC) and low-copy (LC) number plasmids as well as integrated into the genome in a single copy. (e) Reporter GFP was compared to an essential functional tagged-protein controlled by the same promoter (for experimental details see supplementary material).

The spectrum of our experiments spans an array of "control parameters" $\vec{p}$ which covers many of the essential processes affecting protein content in cells. The two organisms used, *E. coli* and *S. cerevisiae*, are distinct in almost every aspect of their cell biology and life style, from gene regulation and expression to cell division and physical characteristics such as shape and volume. A comparative experiment in which some control parameter was varied will reveal the sensitivity of the distribution to that particular parameter. If there is no sensitivity and the distributions are the same, then they do not convey information about that parameter and the two experiments exhibit universal behavior. Given the differences between the organisms, the various regulatory systems and the different experimental conditions, it was not at all obvious *a-priori* that any universality could be found.

Figure 1 shows a collection of distributions measured in such comparative experiments. Despite the clear differences in scale between the distributions they all show common features: all are skewed, unimodal and exhibit extended exponential-like tails. These general features were previously reported in multiple publications, and different mechanisms were proposed to account for them [11, 13, 20-22]. Some of the distributions displayed in Fig. 1 are very similar to one another: for example Figs. 1a and 1c show two indistinguishable distributions of GFP under the control of LacO promoter on a low copy number plasmid in bacteria, and under the control of the GAL10 promoter integrated into the genome in yeast. Similarly, Fig. 1d depicts identical distributions of a reporter GFP expressed under the GAL10 promoter and an essential metabolic protein tagged with GFP at its N-terminal, both integrated into the genome in yeast. Assuming that these similarities are not coincidental, the possibility arises that there is a universal principle underlying protein distributions in proliferating populations of microorganisms.

---

[1] A constitutive promoter is one in which the rate of gene expression is approximately constant. By contrast, a regulated promoter is sensitive to signals and varies its expression rate accordingly.

To test this possibility, we compared the distributions after normalizing out the obvious differences in absolute scales, which are mostly manifested in their mean and standard deviation. The mean $\mu = \langle x \rangle$ reflects the absolute number of proteins in the cell, the strength of the specific GFP used and its behavior in the different biological contexts. The standard deviation $\sigma = \sqrt{\langle x^2 \rangle - \langle x \rangle^2}$ is strongly affected by the dynamic range of protein content that also depends on the particulars of the biological system.

Figure 2 shows all the distributions of Fig. 1 on a common *x*-axis. For each distribution the axis was normalized by subtracting its mean and dividing by its standard deviation. Remarkably, all distributions collapsed to a single curve over almost 10 standard deviations in scaled fluorescence (*x*-axis) and more than 3 decades in probability density. This presentation reveals the universality of the protein distribution *shape* within the entire array of our experimental conditions: the distribution *f* obeys the scaling form

$$(1) \quad f(x;\vec{p}) = \varphi\left(\frac{x - \mu(\vec{p})}{\sigma(\vec{p})}\right)$$

showing that the dependence on the control parameters $\vec{p}$ enters through the mean and standard deviation. Other forms of scaling do not result in such a collapse (Supplementary Fig. 1). Among several well-known skewed distributions we found that the rescaled data can be well fitted by the Frechet distribution, shown by the black curve in Fig. 2, or by a log-normal distribution. Further information on fitting the data is given in Supplementary Figs. 3, 4. It is emphasized that other fitting functions can possibly describe the data equally well; at this stage these are empirical fittings only.

Normalizing out the first two moments resulted in a universally-shaped distribution with zero mean and unit standard deviation, and discarded information on possible relations between the moments in the original, physical units. Plotting these moments one versus the other, one point for each distribution for both bacteria and yeast (Fig. 3a,b), reveals that the variance defines a curve in the plane with very little scatter, that can be well fit by a quadratic function $y = Ax^2 + Bx + C$. Fig. 3c shows a similar relation for many measurements on yeast populations done by fluorescence microscopy [11].

Further support of this relation between moments is found from transient experiments, in which we use the chemostat to switch medium between inducing and repressing conditions of gene expression while still maintaining an exponentially growing culture. Fig. 4 depicts the distributions of GFP under the control of the GAL10 promoter in a yeast population switched from inducing galactose to repressing glucose medium (Fig. 4a) and LacO promoter in bacteria switched from repressing glucose to inducing lactose medium (Fig. 4b). It is seen that the qualitative features of the distributions are maintained throughout the transient but with a time-varying exponential tail. The insets show that in these experiments, as in the steady states, the variance and mean define a quadratic relation with very little scatter.

Finally, we note the quadratic dependence between variance and mean is exhibited also by published genome-wide measurements [17, 20, 23]. In previous work variation was characterized by the ratio between variance and mean squared ("noise"); this measure is a nonlinear combination of moments and does not provide direct information about the relation between them in the presence of measurement errors. When plotted directly, the data are seen to approximate a quadratic function over a broad dynamic range of measured variables (see Supplementary Figure 5).

The generality of the universal behavior that we have found remains to be characterized in further experiments and organisms. Clearly it does not necessarily apply to every biological realization; for example experiments have shown that under some conditions the number of lac permeases in bacteria exhibits a bimodal distribution ([24]; the same group later concluded from a genome-wide study that such distributions are rare [20]). However, an observation of fundamental importance here is the existence of a universality class in biology. The fact that populations of two distinct microorganisms in a broad range of biological contexts exhibit protein distributions that can be scaled by mean and standard deviation to a universal curve is highly significant. The entailed conclusion is that the shape of these distributions cannot convey information on specific biological molecular or cellular mechanisms related to any of the control parameters covered by our experimental conditions. Together with the observed relation between the variance and mean $\sigma = \sigma(\mu)$, (and regardless of its precise functional form), these results imply that by measuring a single variable, e.g. the mean, the entire protein distribution can be reconstructed:

$$(2) \qquad f(x;\vec{p}) = \varphi\left(\frac{x - \mu(\vec{p})}{\sigma(\mu(\vec{p}))}\right) = f(x;\mu).$$

If protein distributions do not reflect any single dominant molecular or cellular mechanism, they must be the integrated outcome of a large number of stochastic events. The masking of individual stochastic events by an integration of many of them is well known in the case of the central limit theorem. Our data, however, exhibit a universal non-Gaussian skewed exponentially-tailed distribution, implying that if a similar "law of large numbers" exists then some of the conditions of the central limit theorem are not fulfilled. What can one say from the data about the possible candidates of these unfulfilled conditions?

The resemblance of the universal curve to the log-normal distribution immediately raises the possibility of a multiplicative process: if cellular protein content was the product of a large number of independent random variables then its distribution would universally converge to a lognormal distribution

$$(3) \qquad L(x;m,s) = \frac{1}{x\sqrt{2\pi s^2}} e^{\frac{(\ln x - m)^2}{2s^2}}$$

Where $m = \langle \ln X \rangle$ and $s^2 = \text{var}(\ln X)$. This function exhibits a scaling behavior in the variable $\ln x$; because of the nonlinearity of the logarithmic transformation, subtracting the mean and dividing by the standard deviation of the measured data would generally not result in a collapse of distributions. Moreover the variance and mean are generally two independent parameters for a collection of lognormal distributions[2]. Therefore this possibility is inconsistent with the data shown in Figures 2 and 3.

A second possibility is that the fluctuations reflect a sum of many random variables which are not independent but rather strongly correlated. This is plausible from a biological point of view, since the different processes that contribute to the protein content of a cell are indeed strongly correlated and reflect different aspects of the same cell's individuality. Moreover considering the protein content as being accumulated over time by many random events, these events are temporally correlated; this can be deduced from single-cell measurements of phenotypic traits along time that typically show a correlation over a few generations [20, 25]. These arguments support a picture where protein fluctuations arise as an integration of random variables correlated in time as well as constrained by correlations to other variables.

Non-Gaussian universal distributions with qualitatively similar features to those measured here were observed in complex physical systems where global fluctuations were measured. Examples include turbulent flows, magnetization in spin systems and other equilibrium systems near phase transitions as well as non-equilibrium systems [26, 27]. In many of these systems, the universal distributions could be well described by one of three universality classes of extreme value statistics (BHP, also known as Generalized Extreme Value – GEV distributions [28]). Recent theoretical work has illustrated a mapping between the extreme values of a set of independent identically distributed random variables and the sum of non-identically distributed ones [29], showing that they have the same non-Gaussian distribution. This raises also the possibility that the measured fluctuations arise from a sum of random variables that are not identically distributed, and reflect an underlying non-stationary process [30].

Although there is no established theory for the appearance of these distributions, much research has recently been devoted to the understanding of this phenomenon using scaling arguments [26] and models of special cases [29]. Inspired by this line of thought we fitted our data to the GEV distributions and found that it could be best described by the Frechet distribution

$$(4) \quad F(x;k,m,s) = \frac{1}{ks}\left(\frac{x-m}{s}\right)^{-\frac{1}{k}-1} e^{-\left(x-m/s\right)^{-\frac{1}{k}}}.$$

---

[2] There is one special case where a collection of different lognormal distributions show such a collapse, and that is when each distribution results from a product of random variables with a different mean, but all have exactly the same variance. This is a highly unlikely assumption in relation to the data and implies a "conspiracy" which is contrary to the observed robustness of the scaling.

While some individual experiments could be better described by a lognormal or Gamma distribution, the pooled dataset was significantly better fit by the Frechet distribution than any other function we have tried (supplementary Figures 3,4). More importantly, it can be shown that a family of Frechet distributions with fixed shape parameter $k$ exhibits both properties of the data – scaling by the first two moments and quadratic dependence of variance on mean (supplementary analysis). Thus, while it may be possible to choose parameters where the lognormal and Frechet distributions are practically indistinguishable over the finite range of measurements, their scaling and symmetry properties are different and only the latter are consistent with the data. In spite of these consistencies, we still regard the fit to a Frechet distribution as a phenomenological description of the data. In the absence of a theory, it is not possible to exclude at this point that other distributions may describe the data equally well. Recent work has illustrated that much ambiguity can occur when inferring the details of a stochastic process from the phenomenology of its statistical properties [30],

The analogy between a cell population and the above mentioned physical systems is still suggestive at this time. However our results call for understanding of the observed universality and for connecting it with other physical systems exhibiting a similar behavior. The connection is not straightforward; a population of cells is not a statistical ensemble of separate realizations as they exhibit long-term correlations [11, 25] and collective effects in gene expression [31]. Searching for such a connection marks a challenging direction for future research on the interface between biology and the physics of complex systems.

**Acknowledgements:** This work was supported in part by a US-Israel Binational Science foundation grant and by the Israel Science Foundation (grant # 496/10, EB). We are grateful to Stefano Ruffo for suggesting a BHP fit to our data.

**Figure Captions**

**Fig. 1: Protein distributions.** GFP fluorescence distributions in populations of bacteria (red) and yeast (black) measured under different conditions by flow cytometry. (a) High and low copy number regulatory promoters in bacteria and yeast: GFP expressed under the LacO promoter on high (circles) and low (squares) copy number plasmids in bacteria, and under the GAL10 promoter on a high copy plasmid (circles) and integrated into the genome (squares) in yeast. (b) Continuous (chemostats; circles) and batch cultures (diamonds) in bacteria and yeast: GFP expressed from high copy number plasmids under the LacO or GAL10 promoters, respectively. (c) Regulated (squares) and constitutive (diamonds) promoters: In bacteria, LacO promoter is compared to ColE1P1 promoter, both on high copy number plasmids, and in yeast the GAL10 promoter is compared to ADH1, both integrated into the genome. All populations were grown in batch cultures. (d) Reporter GFP under GAL10 promoter (squares) and a functional HIS3-GFP N-terminal (stars) and C-terminal (triangles) tagged, both under the GAL1 promoter. All fluorescence levels in this figure were normalized such that the peak of the distributions appears at 1. The probability density is normalized to unit area. Note the logarithmic y-axis.

**Fig. 2: Universal scaling of protein distributions.** (a) All distributions of Fig. 1 are plotted in normalized units by subtracting the mean and dividing by the standard deviation. The symbols represent different experiments as follows: LacO promoter on a high-copy plasmid in bacteria (blue star – grown in chemostat, cyan triangle – in batch); the same on a low copy plasmid in chemostat (red circle); GAL10 promoter on a high copy plasmid in yeast (green triangle – grown in chemostat, magenta squares – in batch); the same integrated into the genome in chemostat (black cross); ADH1 promoter integrated into the genome in yeast grown in batch (blue diamond); ColE1P1 promoter on a high copy plasmid in bacteria grown in batch (red triangle); Fused HIS3-GFP under GAL1 promoter in yeast in chemostat (N-terminal - cyan circles, C-terminal - black pentagram). The black line is the Frechet distribution best fit to the data (see supplementary Figure 4 for details).

**Fig. 3: Relation between mean and variance.** (a) Mean and variance for all experiments in bacteria that appear in Fig. 2 using the same symbols. (b) The same for yeast experiments. (c) Mean and variance for a large collection of experiments on yeast populations, where steady state protein distributions were measured by fluorescence microscopy [11]. GFP was expressed under the control of GAL10 on high or low copy number plasmids, in different background strains grown in chemostat cultures with different dilution rates and limiting nutrients. Fits: $y=Ax^2+Bx+C$ with the parameters: (a) A=0.46, B=-0.39, C=0.1. (b) A=0.17, B=3.1, C=-1.54. (c) A=0.167, B=-0.31, C=-0.015. The data is presented in three panels since fluorescence is not calibrated and therefore different sets of experiments performed under different conditions contain an arbitrary scale factor.

**Fig. 4: Distributions under transient dynamics.** Populations of bacteria (a) and yeast (b) were grown in chemostat to steady state and then switched to different medium. Main figures show protein distribution along time; insets show mean and variance throughout the transient. (a) Bacteria were grown to steady state in glucose and then switched to high concentration of lactose. As a result, an overshoot of induction was seen (blue x; broadest distribution) followed by an adaptive response until a steady state in lactose was reached (green triangles). Then

the culture was switched again to glucose (last two distributions, black crosses and magenta squares). (b) Yeast cells were grown in chemostat to steady state in galactose (broadest distribution; red circles), then switched to glucose. The tail of the distribution gradually decreased following the switch until a very narrow distribution (red stars) was reached. In both experiments the mean and variance displayed a quadratic relation (insets); gray points represents measurements that are not displayed as distributions.

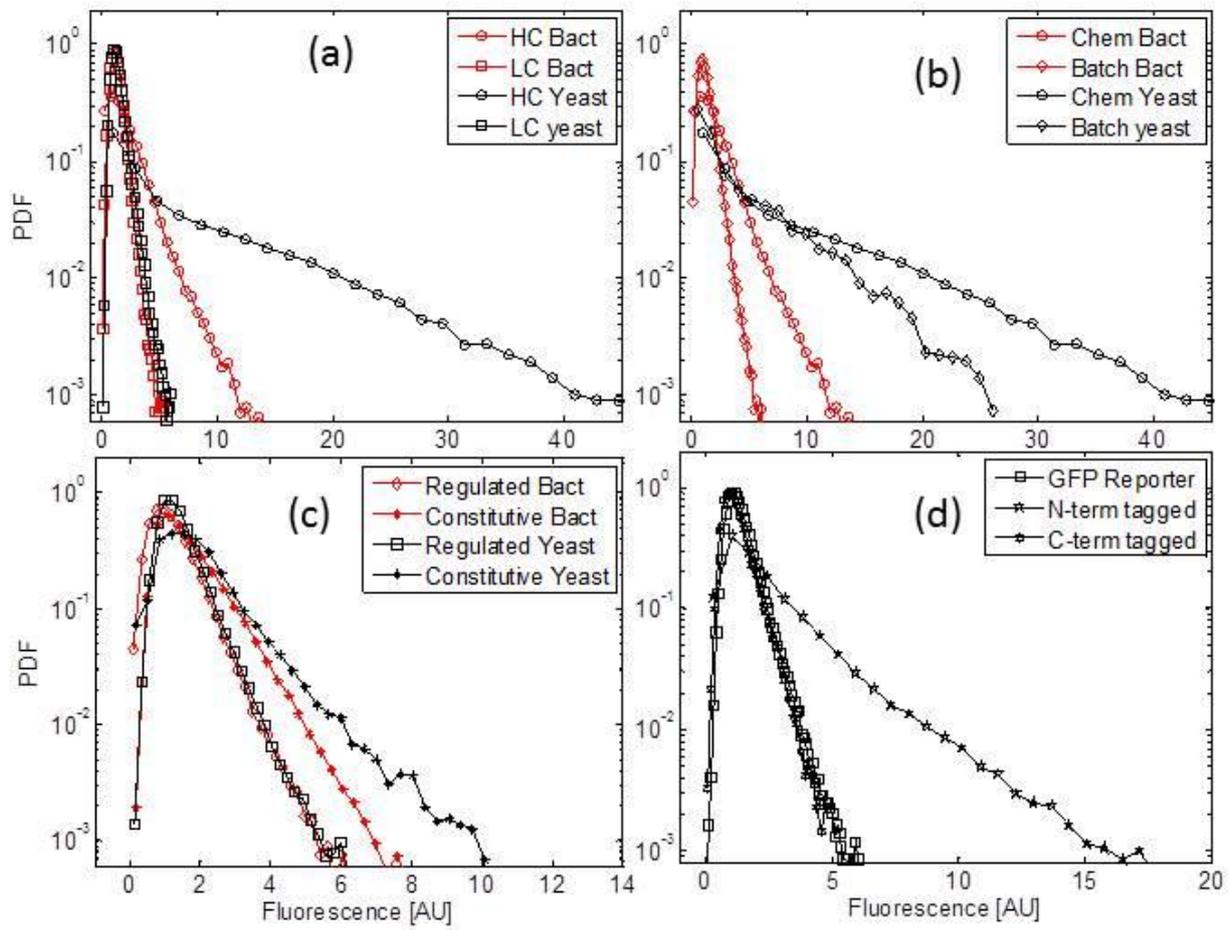

**Figure 1**

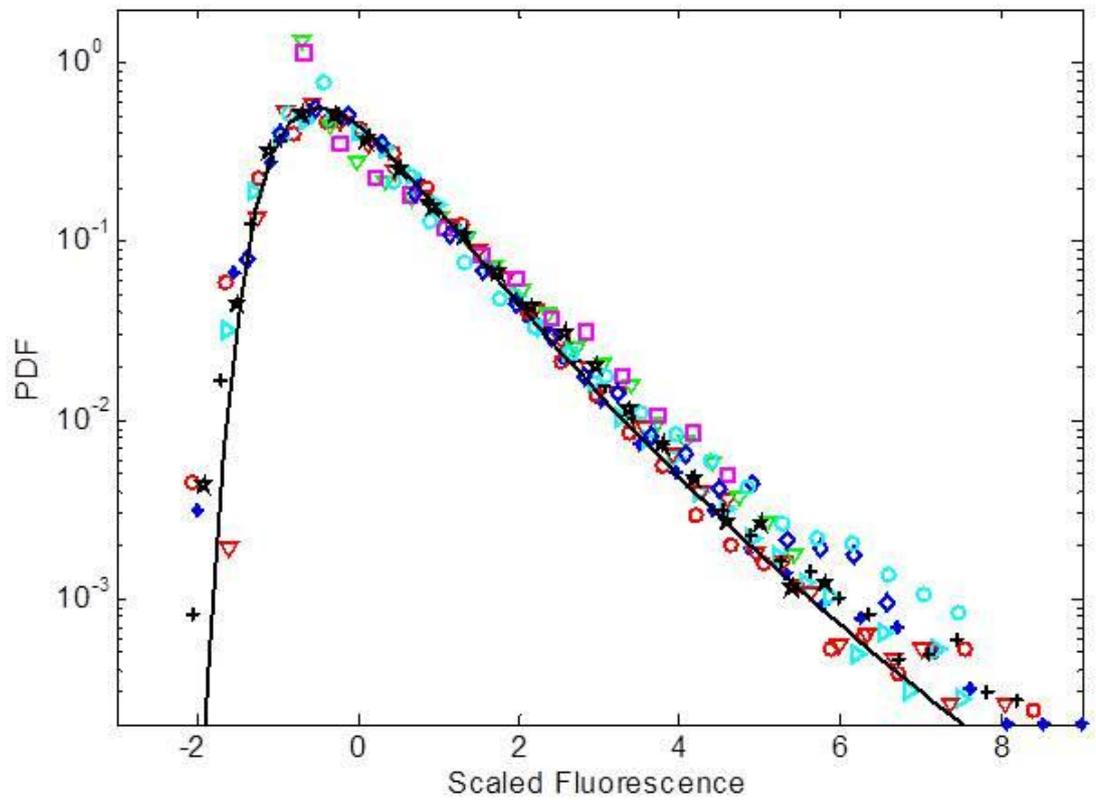

Figure 2

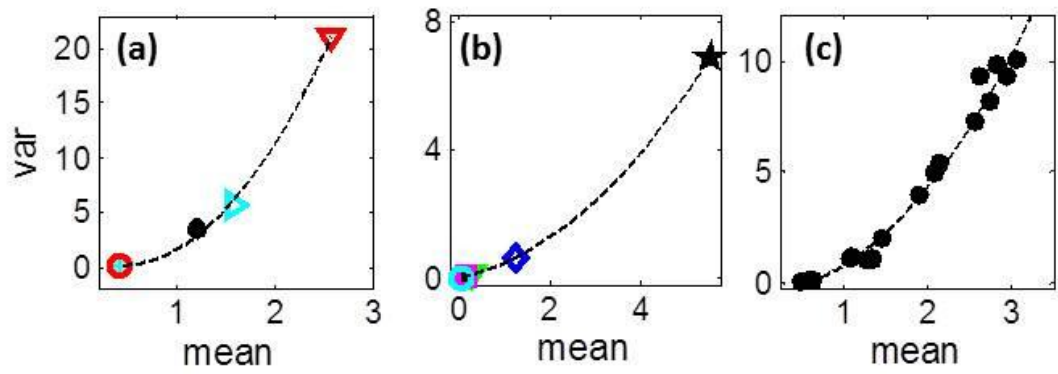

Figure 3

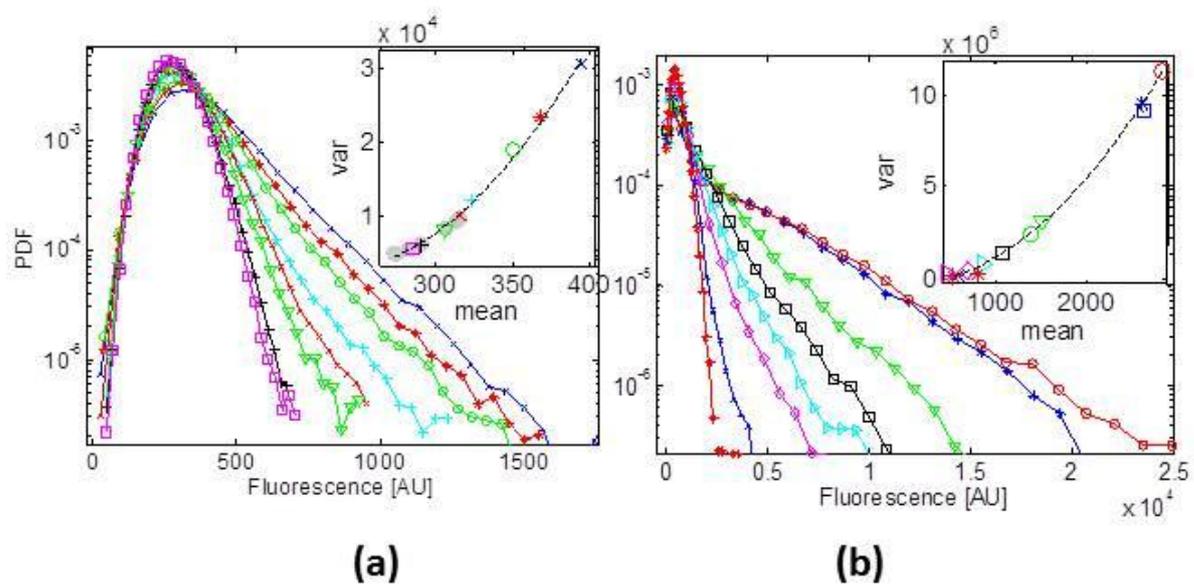

**Figure 4**

# Supplementary material, methods and figures

## I. Experimental methods - Bacteria

*Bacterial Strains and Plasmid:* The gene of the Green Fluorescent Protein (*gfp*) was inserted under the control of the wild type Lac operon promoter (LacO Pr) into either one of two plasmids, the low copy number plasmid (~5) pZS*12wt-GFP, and the high copy number plasmid (~15) pZA12wt-GFP. The wild type MG1655 *E. coli* bacteria were then transfected with these.

*Bacterial Growth Conditions:* Two methods of growing bacteria were used. The continuous culture method (chemostat), in which 50 ml of bacterial culture were grown in the chemostat chamber under steady state conditions. Fresh M9CG (or M9CL) medium (M9 minimal medium supplemented with 0.2g/l casamino acids and 3g/l lactose (or glucose)) was pumped into the growth chamber at the same rate the culture mix was pumped out using a multi-channel peristaltic pump (Ismatic, IDEX Health & Science Group). The jacketed growth chamber was maintained at constant temperature of 32°C by water circulation, and a magnetic stirrer was used to mix the bacterial culture throughout the experiments. In the batch mode cultures, bacteria were grown in 50ml M9CG (or M9CL) medium at 32°C while shaking at 240rpm.

*Flow cytometry*: Measurements of the protein distribution at the different conditions were carried out on 1ml samples that were collected from the bacterial cultures at different times. The fluorescence intensity of GFP at the single-cell level was measured using a flow cytometer (BD LSR II, BD Biosciences).

## II. Experimental methods - Yeast

*Plasmid and strain constructions.* Experiments were carried out with the haploid yeast strain YPH499 [Mata, ura3-52, lys2-801, ade2-101, trp1-Δ63, his3Δ200, leu2Δ1]. Cloning was done by standard methods and was confirmed by fragments analysis and/or by direct sequencing. Transformation was done with the lithium acetate method. *High-copy plasmid:* The plasmid vector pESC-LEU (Stratagene) containing pGAL1-pGAL10 divergent promoter was cloned with GFPS65T under pGAL10 (BglII-gfp-NotI) (*1*). *Single-copy promoter*: A single copy of GFP was integrated into the genome at the LEU2 locus by using the plasmid pRS405 cloned with the GFP downstream the GAL10 promoter. *C-terminal tagged HIS3*: pGAL1-HIS3 tagged with GFP at the C-terminal was constructed in a few steps: First, HIS3 ORF was placed under the pGAL1 in the pESC plasmid containing the diverging GAL promoter described above and then amplified by PCR. The drug resistance gene, Hygromycin B (hphMX) was also amplified. Both PCR amplifications were done with the appropriate primers adding sticky ends to the amplified DNA, allowing homologous recombination to the HO locus (*2*). The results shown in this paper were obtained with a strain containing the insert at the Leu2 locus. To obtain this, both PCR products were combined and inserted to the cell in the same transformation reaction with selection on 300 mg/ml Hygromycin B. In the next step, GFP S65T from the above high-copy plasmid and the drug resistance gene for G418 (kanMX) were amplified by PCR with primers containing proper sticky ends. Both PCR products were combined and inserted to the transformants from the first stage and selected on 200 mg/ml G418. Next, the pGAL1-GFP-HIS3 cassette was amplified from the genomic DNA of the above strain and inserted into plasmid pRS405. The plasmid was linearized by cutting with EcorV at the Leu2 locus and transformed to yeast using selection on a medium lacking leucine and histidine. This transformation leads to recombination of the construct into the Leu2 locus. *N-terminal tagged HIS3*: A high-copy number plasmid pESC-pGAL1-GFP -HIS3 (N terminal) was created in two steps. First HIS3's ORF was inserted to the pESC-Leu2 plasmid under pGAL1. GFP S65T was fused to the N-terminal by amplifying the GFP with primers containing the proper restriction sites. After ligation, the plasmid was transformed into the YPH499 strain using selection on a medium lacking leucine. Integration of pGAL1-GFP-HIS3 (N terminal) into the Leu2 locus was created similarly to the strain with HIS3 tagged at the C-terminal. *Constitutive promoter*: A yeast strain expressing GFP under the constitutive promoter of ADH1 (yEB1102) was prepared in two steps. First, the plasmid p406ADH1 (Addgene plasmid 15974; we thank Fred Cross and Nicolas Buchler for this plasmid) cut with BamHI and EcoRI was cloned with GFP-S65T. Cloning was confirmed by PCR and by fragments analysis. In the second step, this plasmid was cut with StuI and transformed into the YPH499 strain by selection on uracil.

*Chemostat growth*: Cells were grown in a homemade chemostat (*3*) in synthetic dropout medium, with the appropriate amino-acid supplement an galactose as a sole carbon source. Throughout the experiments, the sugar was always in excess. Medium (concentrations in g/l): 1.7 yeast nitrogen base without amino-acids and ammonium sulfate, 5 ammonium sulfate, 1.4 amino-acids dropout powder (without tryptophan, histidine, leucine and uracil; Sigma), with 0.01 l-tryptophan, 0.005 uracil and 2% galactose. Growth in the chemostat was limited by the concentration of the amino acid supplement. The chemostat contains ~130 mL ($10^9$-$10^{10}$ cells at steady state).

*Batch growth*: Cells in batch culture were grown in a similar medium composition to that of the chemostat with 0.04 g/l L-tryptophan, 0.02 g/l uracil and (for strains without HIS3), 0.02 g/l histidine, 0.06 g/liter leucine for the strain with the constitutive promoter, and 2% galactose as the sole carbon source.

*Flow cytometry* measurements were performed using LSR II (Becton Dickinson) with a 488nm excitation laser and a 530/30 emission filter. In every experiment 10,000-30,000 cells were measured.

### III. Analysis of scaling for Frechet distribution

The scaling properties of the Frechet distribution can be analyzed by considering its simple expression for the cumulative distribution function

$$(1) \quad F(x;k,s,x_0) = e^{-\left((x-x_0)/s\right)^{-\frac{1}{k}}}.$$

Its mean and variance are given by

$$(2) \quad \mu = \langle x \rangle = x_0 + s\gamma_1(k)$$
$$\sigma^2 = \text{var}(x) = s^2 \gamma_2^2(k)$$

Where we have defined $\gamma_1(k) = \Gamma(1-k)$, $\gamma_2^2(k) = \Gamma(1-2k) - \left(\Gamma(1-k)^2\right)$ and $\Gamma$ is the Gamma function. The Frechet distribution is a function of a shifted and scaled variable, but not by the first and second moments of the distribution. Defining a new variable by

$$(3) \quad y = \frac{x-\mu}{\sigma}$$

It is seen that the original scaled and shifted variable is a linear function of $y$:

$$(4) \quad \frac{x-x_0}{s} = \frac{s\gamma_2(k)y + x_0 + s\gamma_1 - x_0}{s} = \gamma_2(k)y + \gamma_1(k)$$

And therefore, for fixed $k$, the distribution is also a scaling function of $y$:

$$(4) \quad F(x;k,s,x_0) = F_k\left(\frac{x-x_0}{s}\right) = \varphi_k\left(\frac{x-\mu}{\sigma}\right)$$

proving that a family of Frechet distributions with fixed $k$ obeys the scaling form that we have found in the experiments. To show that the variance is a quadratic function of the mean, substitute $s = \frac{\mu - x_0}{\gamma_1(k)}$ to find

$$(5) \quad \text{var}(x) = s^2 \gamma_2^2(k) = \left(\frac{\mu - x_0}{\gamma_1(k)}\right)^2 \gamma_2^2(k) = a\mu^2 + B\mu + C.$$

It is noted that the other two universal distributions of extreme value statistics, the Gumbel and Weibull distributions, obey the same scaling form and the same quadratic relation between moments. This holds true also for Gamma distributions of a fixed order, but not for log-normal distribution.

# Supplementary Figures

**Figure S1: Lack of universality with different forms of scaling.** The same distributions shown in Figure 2 of the main text are displayed on differently normalized *x*-axes. (a) Fluorescence is normalized to unit mean, namely the *x*-axis represents the fluorescence divided by the mean. (b) Fluorescence is normalized to unit standard deviation. These normalizations could be expected to make sense since the distributions are dominated by exponential-like tails (see also (*4*)). (c) Fluorescence is normalized such that the peak of the distribution is at 1, and plotted on a logarithmic *x*-axis and a linear *y*-axis. This normalization was suggested in previous work where universal features of size distributions was reported (*5*). It is, in fact, the same normalization used in Figure 1 of the main text, but presented on different axes that enhances the small-*x* regime rather than the tail behavior.

**Figure S2: Higher precision of universality in a subset of the data.** In Figure 2 of the main text, experiments from yeast cells with high copy number plasmid display a small peak which deviates from the universal curve. Here we show a subset of five experiments (using the same symbols as in Fig. 2) which display a higher accuracy of the scaling without the deviation at the peak, and still cover a significant range of the conditions we have explored: two measurements out of the five are from bacteria while the other three are from yeast; two are expressed under promoters integrated into the genome while two others are expressed on plasmids (high and low copy); four are under strongly regulated metabolic promoters while one is under a constitutive promoter, ADH1 in yeast; four are reporter proteins while one (fused HIS3-gfp N-terminal) is a functional protein tagged by GFP. All populations were grown in continuous culture but with different growth rates (dilution times). Also shown in the figure is a fit to the Frechet distribution. Note the high accuracy of the fit over a wide range of scaled fluorescence (10 standard deviations) and of probability (3 decades).

**Figure S3: Fits of individual measurements**. We applied fitting of each experiments to three skewed distributions: Gamma, Frechet and log-normal. These distributions are usually defined for non-negative variables. In relation to the scaled distributions presented in Fig. 2, this implies that an additional fitting parameter $x_0$, a shift on the x-axis which defines the variable to be in the range $x > x_0$. Thus each of the distributions has 3 parameters:

$$G(x;a,s,x_0) = \frac{1}{s^a \Gamma(a)} (x-x_0)^{a-1} e^{-(x-x_0)/s}$$

$$F(x;k,s,x_0) = \frac{1}{ks}\left(\frac{x-x_0}{s}\right)^{-\frac{1}{k}-1} e^{-\left((x-x_0)/s\right)^{-\frac{1}{k}}}$$

$$L(x;m,s,x_0) = \frac{1}{(x-x_0)\sqrt{2\pi s^2}} e^{-\frac{(\ln(x-x_0)-m)^2}{2s^2}}$$

.

All data are shown using the same symbols as in Fig. 2 of the main text. It is seen that some of the experiments are best described by the Frechet distribution while others by a lognormal distributions (the parameter R in the legend represents the absolute area of the error in the cumulative distribution between the data and the fit). The two yeast experiments which show a high peak near the origin are better described by Gamma distributions. We emphasize that the fits are only descriptive, and at this stage we cannot attach any particular explanation to any one of them. Also to be noted is the fact that, for example, the Frechet and lognormal distribution are hardly distinguishable over the range of measurement for some sets of parameters (for discussion of the lognormal distribution, its many possible sources and its similarity to other distributions, see (*6, 7*)). The log-normal distribution can arise when there is an underlying Gaussian distribution and a nonlinear logarithmic transformation (see, for example, model proposed in (*8*)), or as a limiting universal distribution for a multiplicative process. The universal behavior we observe is inconsistent with a specific logarithmic transformation on any particular variable; the second possibility of a universal multiplicative process is inconsistent with the scaling behavior (see discussion in main text).

**Figure S4: Fit of all rescaled data pooled**. To within the accuracy of the scaling, it is of interest to pool together all rescaled data and test the different fits. It is seen that the Frechet distribution provides an excellent fit for almost all dynamic range, except for some discrepancies near the peak which are affected by the two yeast experiments mentioned in Fig. S3. This is the same function plotted in Fig. 2 of the main text with parameters: *$x_0$=7.5, s=7.1, k=0.095*. The log-normal distribution also gives a good fit (with a similar discrepancy near the peak) and parameters *$x_0$=-0.44, m=0.74, s=0.4*

**Figure S5: Relation between variance and mean in published genome-wide data.** Means and standard deviations computed for each of 1018 genes in *E. coli* were taken from Supplementary online material of (*10*). **(a)** Variance as a function of mean on double logarithmic axes shows a clear quadratic dependence in the range of ~5-$10^4$ proteins per cell on average, which defines over 99% of the dynamic range in measured protein abundance. The quadratic dependence is exhibited by approximately 2/3 of the data points; for rest, rare proteins with abundance <5, the dependence is consistent with a linear dependence of variance on mean.

**(b)** Same data points showing $(CV)^2$=(variance)/(mean$^2$) as a function of mean, reproducing Figure 2B of Ref. (*10*).

**Figure S1**

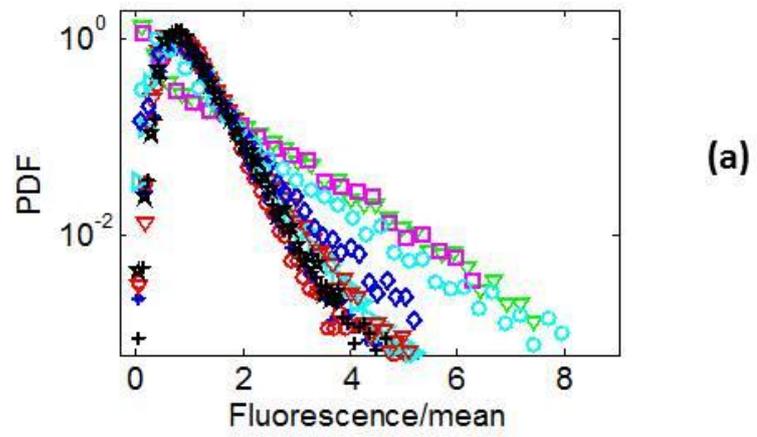

(a)

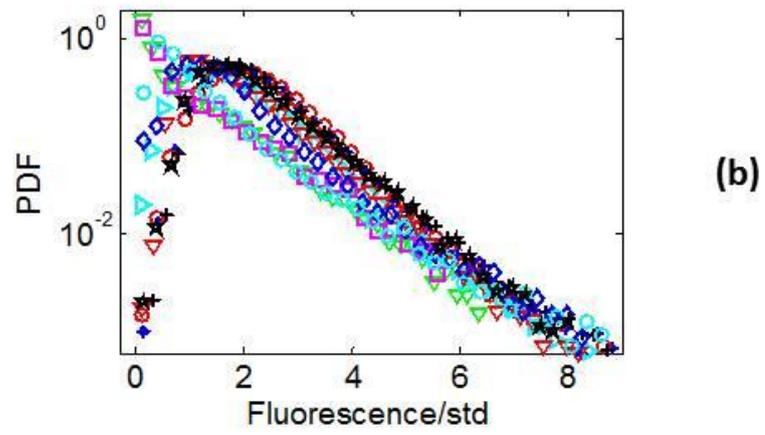

(b)

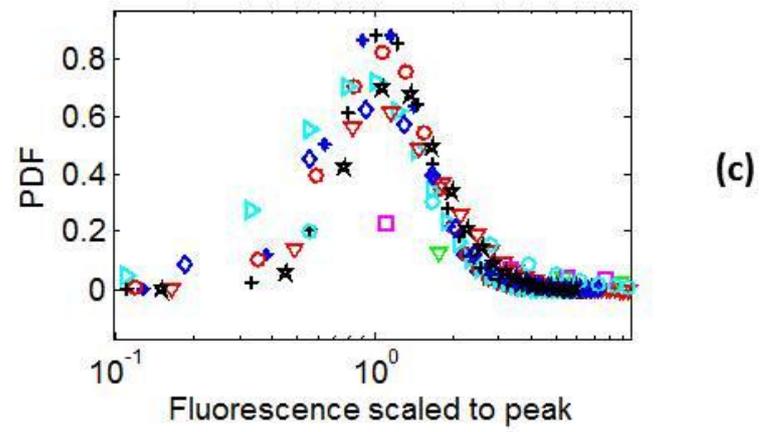

(c)

**Figure S2**

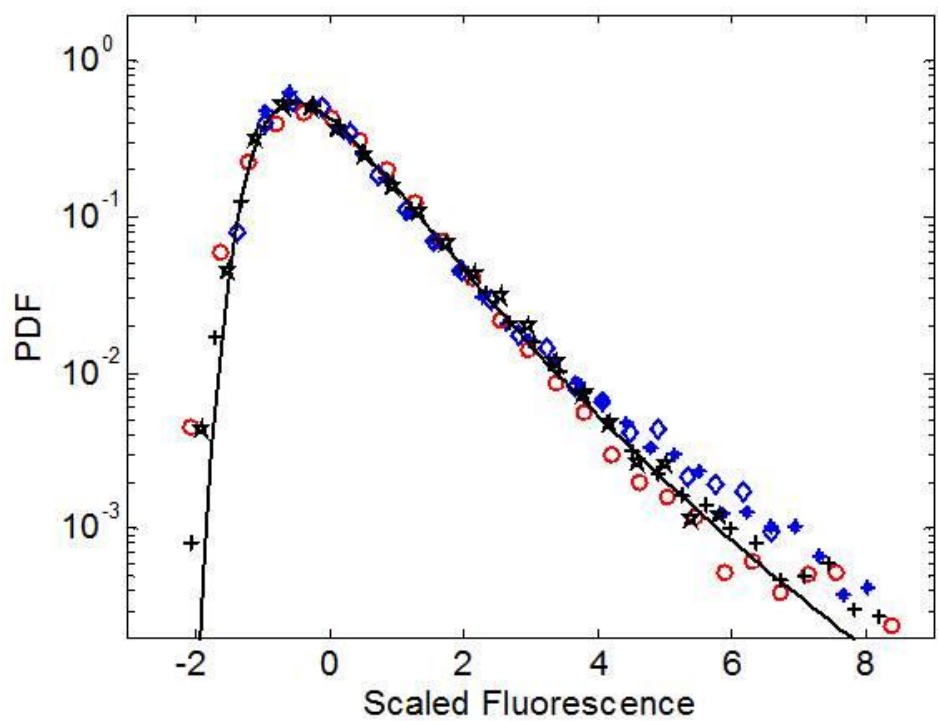

## Figure S3 : Bacteria

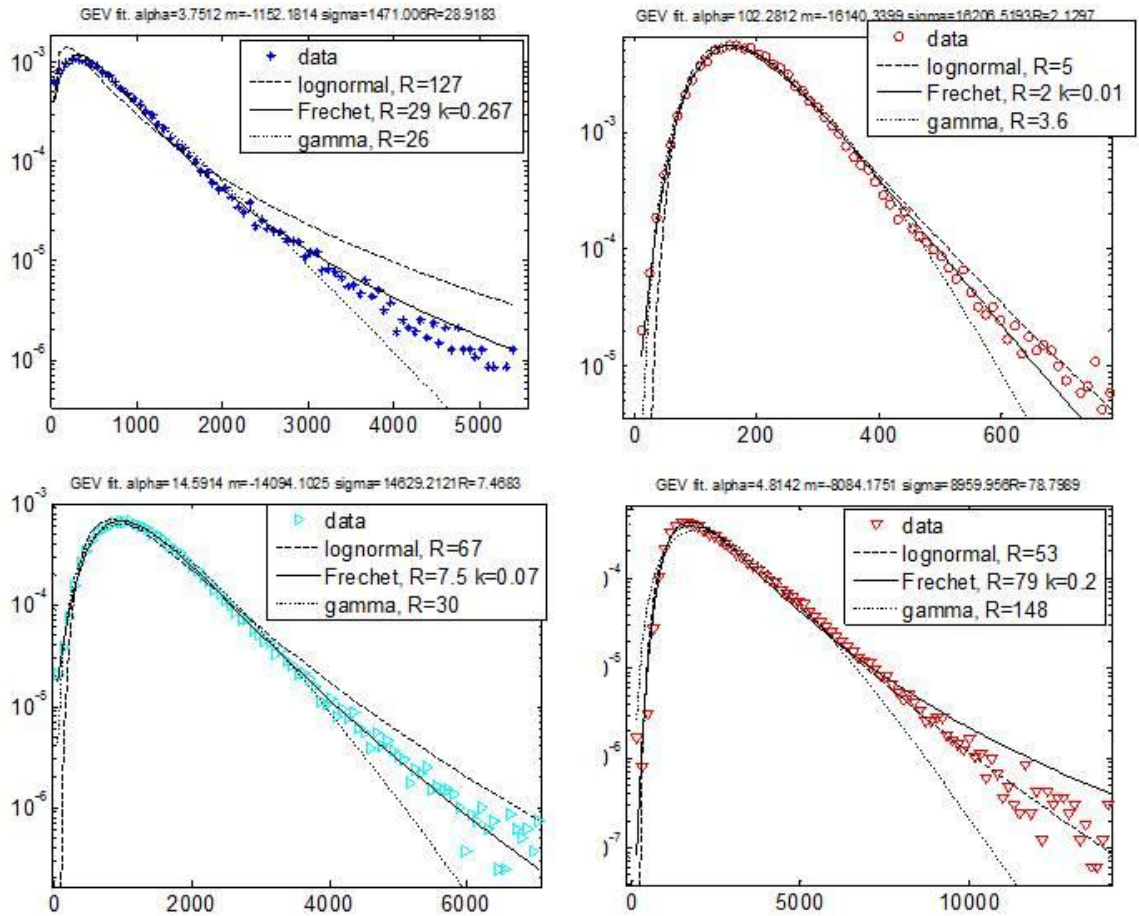

# Figure S3: Yeast

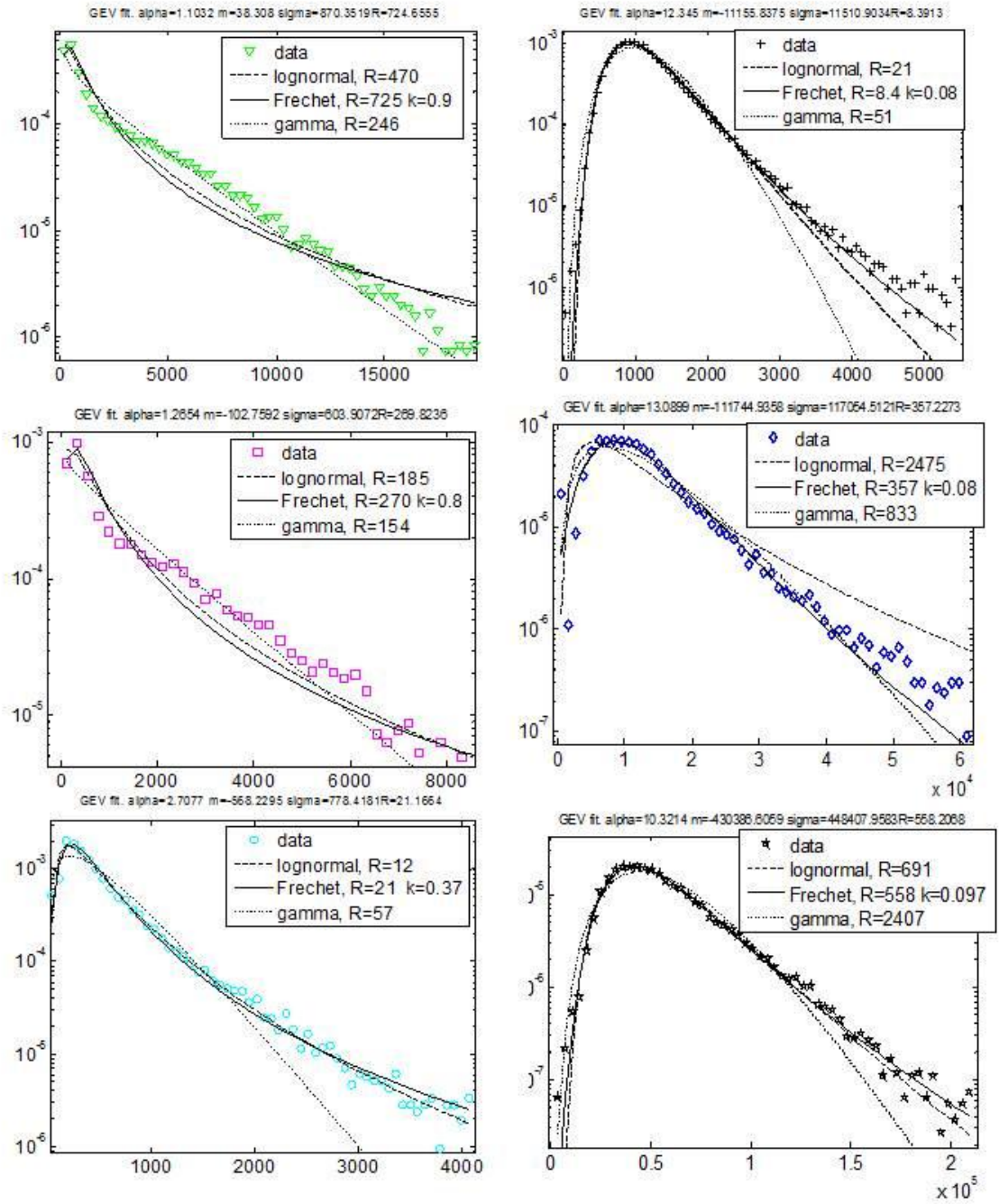

**Figure S4**

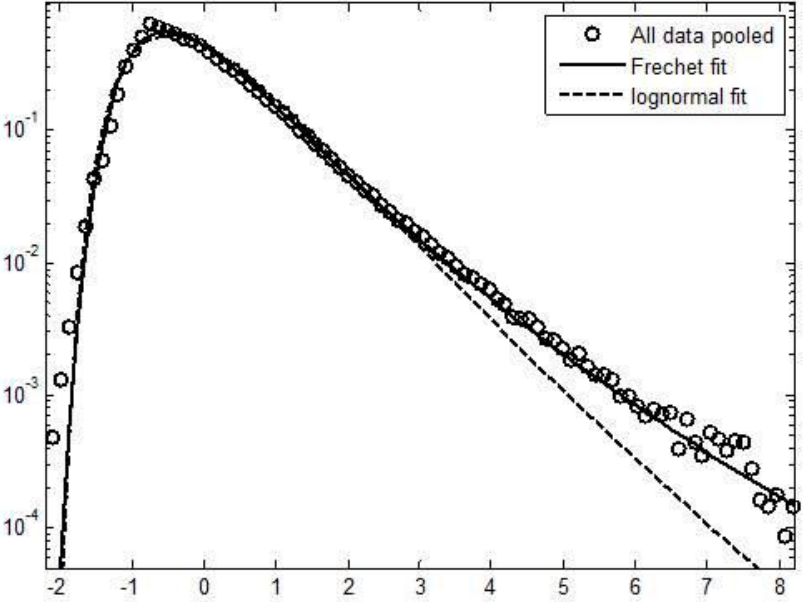

**Figure S5**

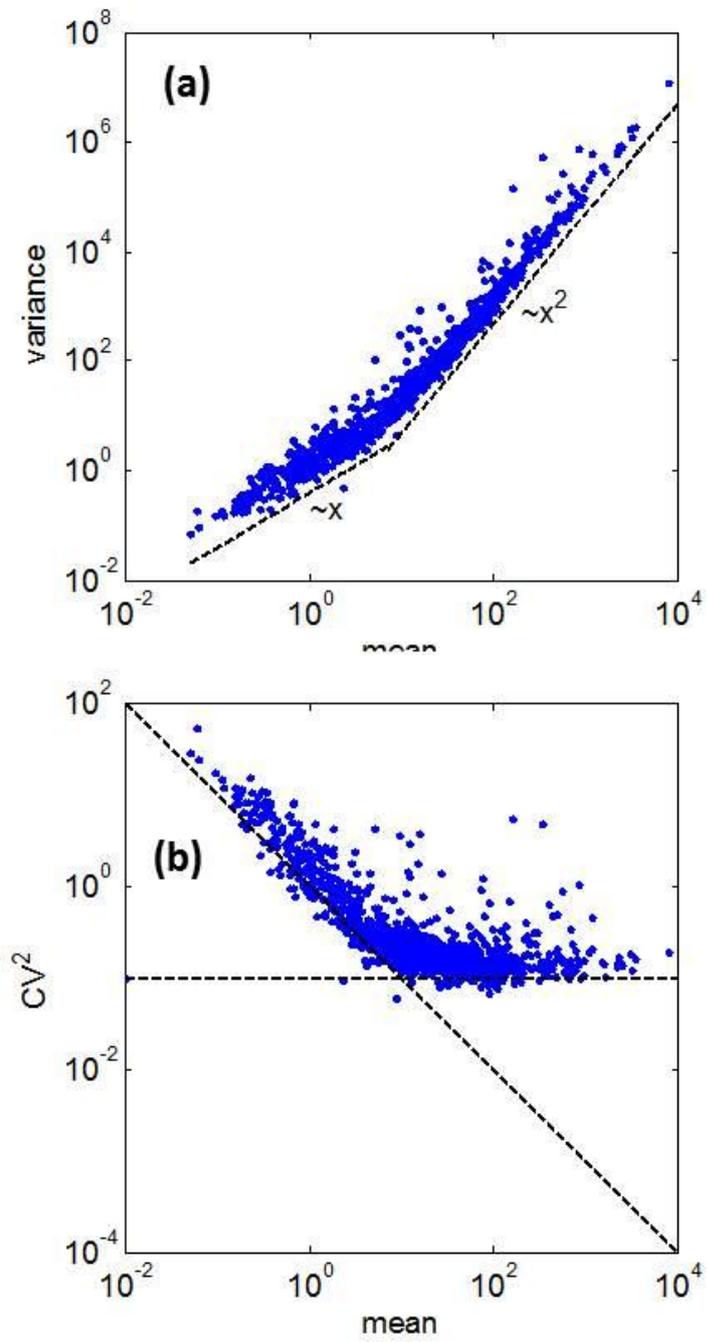